\documentclass[aps,prd,onecolumn,preprint,groupedaddress,showpacs,nofootinbib]{revtex4}

\usepackage{graphicx}
\usepackage{dcolumn}
\usepackage{bm}

\usepackage{color}
\usepackage{bm}

\definecolor{darkred}{rgb}{.8,0,0}

\definecolor{darkblue}{rgb}{0,0,.7}

\definecolor{darkgreen}{rgb}{0,.8,0}

\usepackage{amsmath, amsthm}
\def\ben{\begin{enumerate}} \def\een{\end{enumerate}}
\def\beq{\begin{equation}} \def\eeq{\end{equation}}
\def\beqn{\begin{equation*}} \def\eeqn{\end{equation*}}
\def\bea{\begin{eqnarray}} \def\eea{\end{eqnarray}}
\def\ba{\begin{array}} \def\ea{\end{array}}
\def\beann{\begin{eqnarray*}} \def\eeann{\end{eqnarray*}}
\def\beasn{\begin{sneqnarray}} \def\eeasn{\end{sneqnarray}}

\begin{document}

\author{Naresh Dadhich}
\thanks{email: {\tt nkd@iucaa.ernet.in}}
\affiliation{Centre for Theoretical Physics, Jamia Millia Islamia, New Delhi~110025, India, and 
Inter-University Centre for Astronomy and Astrophysics, Post Bag 4, Pune 411 007, 
India}
\author{Josep M Pons}
\thanks{E-mail:\tt{pons@ecm.ub.edu}}
\affiliation{Departament d'Estructura i Constituents de la Mat\`eria and 
Institut de Ci\`encies del Cosmos (ICCUB), Facultat de Física, Universitat de Barcelona, Diagonal 647, 
E-08028 Barcelona, Catalonia, Spain.}

\pacs{}

\title{On static black holes solutions in Einstein and Einstein-Gauss-Bonnet gravity with 
topology ${\bf SO(n) \times SO(n)}$}

\begin{abstract}
We study static black hole solutions in Einstein and Einstein-Gauss-Bonnet gravity with product 
two-spheres topology, ${\bf SO(n) \times SO(n)}$, in higher dimensions. There is 
an unusual new feature of Gauss-Bonnet black hole that the avoidance of non-central naked singularity 
prescribes a mass range for black hole in terms of $\Lambda>0$. For Einstein-Gauss-Bonnet black hole a 
limited window of negative values for $\Lambda$ is also permitted. This topology encompasses black string 
and brane as well as a generalized Nariai metric. We also give new solutions with product two-spheres of 
constant curvatures.
\end{abstract}

\maketitle
\flushbottom

\section{Introduction}

The study of gravity in higher dimensions was given great impetus by string theory for which it is a 
natural framework. One of the most compelling pictures that emerges is that all matter fields are 
believed to remain 
confined to the usual $4$-dimensional spacetime - $3$-brane, while gravity can propagate in higher 
dimensions. At the root of this perception, we believe is the unique gravitational property of universality 
- its linkage to all that physically exist. On the other hand it has also been argued by one of us 
\cite{d0, d1} purely on classical considerations that gravity cannot be entirely confined to a given 
dimension. High energy effects would ask for inclusion of higher orders in Riemann curvature in the action, 
then if the resulting equation is to remain second order, it has to be the Lovelock polynomial which makes 
non-trivial contributions only in $D>4$. That is, high energy effects could only be realized in higher 
dimension \cite{d1, d2} (and the  references therein). It is envisioned as a general guiding principle that anything universal should not 
be constrained from outside but should rather be left to itself to determine its own playground. This is 
precisely what Einstein gravity does. Since it is universal and hence it can only be described by spacetime 
curvature \cite{d0} and it is that which determines the gravitational law \cite{d1}. It is important to 
note that it is not prescribed from outside like the Newtonian law. Similarly higher dimension should 
also be dictated by some property of gravity like the high energy effects. Apart from  a  strong suggestion, we 
have not yet been able to identify a gravitational property that asks for higher dimension. In the absence 
of such a guiding direction, it is a prudent strategy to probe gravitational dynamics in higher dimensions 
so as to gain deeper insight. We believe this is the main motivation for higher dimensional investigations 
of gravity. 

\vspace{4mm}

The first question that arises is, what equation should describe gravitational dynamics in higher 
dimensions, should it be Einstein  or should it be its natural generalization Lovelock equation? 
Einstein equation is linear in Riemann while Lovelock is a homogeneous polynomial yet it has remarkable 
property that resulting equation still remains second order quasilinear. The higher orders in Riemann 
become relevant only in higher dimensions. If for physical reasons like high energy effects, higher orders 
in Riemann are required to be included and the equation should continue to remain second order, Lovelock 
equation is uniquely singled out and that then requires higher dimensions for realization of higher order 
Riemann contributions \cite{d0, d2}. The next question is, should the equation be Einstein-lovelock 
(for a given order $N$, all terms $<N$ are also included) or pure Lovelock (only one $N$th order term plus 
the cosmological constant which is the $0$th order term)? 
It has been shown that pure Lovelock equation has the unique distinguishing property that vacuum for 
static spacetime in all odd $D=2N+1$ dimension is kinematic meaning vanishing of $N$th order Ricci 
implies the corresponding Riemann zero \cite{d3, d4, d5}. For $N=1$ Einstein gravity, it is kinematic 
in $D=3$, and becomes dynamic in next even dimension $D=4$. For pure Lovelock, this is the unique 
feature for odd $D=2N+1$ and even $D=2N+2$ dimensions. What order of $N$ should be involved in 
gravitational dynamics is determined by dimension of spacetime, for example, $D=3, 4$ it is $N=1$ 
Einstein, $D=5, 6$, it is $N=2$ Gauss-Bonnet, and so on. 

\vspace{4mm}

Static vacuum solutions are the simplest and most effective tools for probing a new gravitational 
setting. 
Beginning with E-GB black hole solution by Boulware and Deser \cite{BD} and independently by Wheeler 
\cite{Wh}, and its generalization to general Lovelock \cite{Wh, Whit, BTZ-dc, DPK}, all these black 
holes had horizon with spherical topology having constant curvature. The next order of generalization 
was to seek more general horizon topology of product of maximally symmetric spaces for Einstein 
space horizon. Note that  product space no longer remains  maximally symmetric, however  its Riemann 
curvature is covariantly constant\footnote{ Though in the literature, following Ref. \cite{Dotti1}  
this product space is termed 
non-constant curvature space, its Riemann curvature is indeed covariantly constant but not maximally 
symmetric.}.  For Einstein space, note that $W_{abcd;e} = R_{abcd;e}$, and hence now Weyl curvature is 
covarianly constant which for maximally symmetric Riemann is zero. We would therefore term the product 
space horizon as Weyl constant space. The first interesting solution with this generalization was 
obtained for E-GB black hole by Dotti and 
Gleiser \cite{Dotti1} with horizon space being Weyl constant Einstein space, as realized by  product of two spheres. This  is the case we will concern ourselves in this paper. The measure of Weyl-constancy is expressed as the square of Weyl curvature which makes a non-trivial 
contribution to gravitational potential of the hole. In this paper we shall discuss solutions for 
which  horizon is a product of two spheres. It turns out that for Einstein-Hilbert case for 
gravitational potential of the hole, neither does it matter whether two spheres are of equal curvature 
(and dimensionality) or not, nor does whether 
topology is of one or two spheres. On the other hand, for GB case, they have to be of equal curvature, 
and dimensionality. This is because in the latter case Riemann  tensor is directly 
involved in the equation while in former it is only Ricci tensor, and that is why the latter is more 
restrictive. In 
other words, for Einstein black hole, horizon space need not even be an Einstein space while for GB and 
higher order 
Lovelock it has always to be an Einstein space. There has been a spurt of activity in studying various 
aspects of Dotti-Gleiser black hole in terms of its uniqueness, thermodynamics and stability by various 
authors \cite{Zegers, Dotti2, Maeda}.   

\vspace{4mm}

There is yet another motivation for this paper. The study of spaces with some rotational symmetries 
in general relativity has been strongly motivated by the property that it provides a rich spectrum of 
different phases of black objects with distinct topologies for horizons, see for a review 
\cite{Emparan:2008eg}. The simplest realization of it is provided by the usual $4$-dimensional 
Schwarzschild black hole with an extra dimension of radius $L$ added. If the Schwarzschild radius is 
much smaller than radius of extra dimension, it would 
resemble to $5$-dimensional Schwarzschild black hole with horizon topology $S^3$. On the other hand if 
the black hole radius is bigger than extra dimension radius, it would describe a black string with 
horizon topology $ S^2 \times S^1 $. This means there does occur a local change in horizon topology 
as radius of hole increases. It turns out that this change is brought about \cite{Kol:2002xz}, see also 
\cite{Emparan:2011ve}, through a cone over $S^2\times S^2$ (for brevity, we shall term it two-spheres 
topology) by seeking a Ricci flat metric for the cone. This construction could as well be looked upon as 
solid angle deficit for each $S^2$. Note that angle deficit describes a cosmic string for which the 
Riemann curvature vanishes 
while solid angle deficit for which the Riemann curvature is non-zero describes a global 
monopole \cite{bv}. The interesting question that arises is whether contribution of solid angle 
deficit of one sphere exactly cancels out that of the other giving rise to Ricci flat space. It is 
remarkable that this is precisely what happens for Dotti-Gleiser black hole \cite{Dotti1} with 
two-spheres topology. The two-spheres topology harbours a static black hole with constant Ricci 
(Einstein space) but non constant Riemann curvature horizon. 

\vspace{4mm}

In this paper, we would like to study the more general case of  $S^{d_1}\times S^{d_2}$ for 
$\Lambda$-vacuum solutions of Einstein, Gauss-Bonnet (GB) and Einstein-Gauss-Bonnet (E-GB) equation. 
That is,  a 
($d=d_1+d_2+2$)-dimensional spacetime harbours a static black hole with two-spheres topology 
$S^{d_1} \times S^{d_2}$ for Einstein and $S^{d_0} \times S^{d_0}$ with $d_1=d_2=d_0$ for GB and 
E-GB gravity.  
In the latter case we show that horizon space $S^{d_0} \times S^{d_0}$ has constant Weyl curvature. One of the new features of these GB and E-GB black holes is that 
there can occur a non-central naked singularity which could however be avoided by prescribing a range 
for black hole mass in terms of a given $\Lambda$. It is noteworthy that presence of positive $\Lambda$ 
is therefore necessary for existence of these black holes for GB gravity. That is, $\Lambda$ plays a very 
critical role in this setting as is the case 
for stability of pure Lovelock black hole where it renders otherwise unstable black hole stable \cite{ganno}.  

\vspace{4mm}

Note that we obtain the solutions using a technique different from that of \cite{Dotti1}. Starting from 
the action principle we proceed in two steps. First we perform a consistent truncation together with a 
dimensional 
reduction of the Lagrangian, ending up with a reduced Lagrangian with four mechanical degrees of freedom 
(instead of field theoretical ones). This Lagrangian describes static metrics with 
${\bf SO(n) \times SO(n)}$ 
symmetry. In the second step we introduce an ansatz concerning the radial variable for the spheres and 
the problem becomes an equation for a single degree of freedom. 
Then, as is the case for general Lovelock vacuum equation in spherical symmetry, it all reduces to an 
algebraic equation.  

\vspace{4mm}

The paper is organized as follows: We begin with Einstein black hole and set up the general framework of 
consistent truncation for solving gravitational equations. We thus obtain black hole solutions by this 
alternative method. It is then followed by the general setting of black hole solutions with horizon 
consisting of product spaces of constant curvature. We study various physical features of these black holes 
including prescription of allowed mass range, thermodynamics and thermodynamical stability. 
Next we use the same truncation technique to find solutions of Einstein, GB and E-GB with two spheres 
of constant curvature, and obtain the generalized Nariai metric \cite{nar}.
We end with a discussion.

\section{Einstein black hole}

We begin with the general spherically symmetric metric with two-spheres topology 
$R^2\times S^{d_1}\times S^{d_2}$ which is written as follows: 
\begin{equation}
ds^2 = -A(r)\,dt^2 + B(r)\,dr^2 + C(r)\,d S^2_{(d_1)}+ D(r)\,d S^2_{(d_2)}\,,
\label{construnc}
\end{equation}
in $d=d_1+d_2+2$ dimensions. We use the notation of indices $(0,1)$ for $(t,r)$, $(a,b,...)$ for 
the angular coordinates of the first sphere $S^{d_1}$ and $(a',b',...)$ for those of the second 
sphere $S^{d_2}$. Keeping the four functions $A(r), B(r), C(r), D(r)$ as the unkown variables is a 
consistent truncation ansatz, which means that the direct substitution of (\ref{construnc}) into the 
Lagrangian will give the same equations of motion (EOM) (from the truncated Lagrangian) as if directly 
substituted into the EOM of the original Lagrangian (see details concerning consistent truncations in 
\cite{Pons:2003ka,Pons:2006vz}).

Under this generic ansatz, the only nonvanishing components of the Riemman tensor are:
\begin{eqnarray}
{R_{01}}^{01}&=& \frac{A(r) A'(r) B'(r)+B(r) \Big(A'(r)^2-2 A(r)
   A''(r)\Big)}{4 A(r)^2 B(r)^2} =:L(0,1)\,,\nonumber\\
{R_{0a}}^{0a}&=&-\frac{A'(r) C'(r)}{4 A(r) B(r) C(r)}=:L(0,a)\,,\nonumber\\
{R_{1a}}^{1a}&=&\frac{C(r) B'(r) C'(r)-2 B(r) C(r) C''(r)+B(r)
   C'(r)^2}{4 B(r)^2 C(r)^2}=:L(1,a)\,,\nonumber\\
\end{eqnarray} \begin{eqnarray}  
{R_{ab}}^{ab}&=&\frac{1}{C(r)}-\frac{C'(r)^2}{4 B(r) C(r)^2}=:L(a,b)\,,\quad  (a\neq b)\nonumber\\
{R_{0a'}}^{0a'}&=&-\frac{A'(r) D'(r)}{4 A(r) B(r) D(r)}=:L(0,a')\,,\nonumber\\
{R_{1a'}}^{1a'}&=&\frac{D(r) B'(r) D'(r)-2 B(r) D(r) D''(r)+B(r)
   D'(r)^2}{4 B(r)^2 D(r)^2}=:L(1,a')\,,\nonumber\\
{R_{a'b'}}^{a'b'}&=&\frac{1}{D(r)}-\frac{D'(r)^2}{4 B(r) D(r)^2}=:L(a',b')\,,\quad  (a'\neq b')
\,,\nonumber\\
{R_{a a'}}^{a a'}&=&-\frac{C'(r) D'(r)}{4 B(r) C(r) D(r)}=:L(a,a')\,.
\label{riem}
\end{eqnarray}
The Einstein Lagrangian $\sqrt{-g}( R - 2 \Lambda)$, for the metric (\ref{construnc}) reads as follows:
\begin{eqnarray}
L_{EH}&=&\sqrt{-g}\Big(2\, L(0,1)+ 2\, d_1\Big(L(0,a)+L(1,a)\Big)+2\, d_2\Big(L(0,a')+L(1,a')\Big)
\nonumber\\
&+& d_1(d_1-1)L(a,b) + d_2(d_2-1)L(a',b')+ 2\, d_1d_2 L(a,a')- 2 \Lambda\Big)\,.
\label{leh}
\end{eqnarray}
where the density factor $\sqrt{-g}$ is (up to the volume of the spheres, here irrelevant)
$$
\sqrt{-g}\to\sqrt{A(r) B(r)} C(r)^{\frac{d_1}{2}}D(r)^{\frac{d_2}{2}}\,
$$
It is well known that null energy condition as well as the fact that the radial photon experiences no
 acceleration \cite{ein-new} require $B(r)=\frac{1}{A(r)}$, and we set $C(r) =\frac{ r^2}{ k_1}, 
D(r)=\frac{ r^2}{k_2}$ where $k_1, k_2$ are constants.

The metric thus takes the form 
\begin{equation}
ds^2 = -A(r)\,dt^2 + \frac{1}{A(r)} dr^2 + \frac{ r^2}{ k_1}\,d S^2_{(d_1)}
+\frac{ r^2}{ k_2}\,d S^2_{(d_2)}\,,
\label{construncnext}
\end{equation}
The constants, $k_i$ are fixed as $k_i=\frac{d-3}{d_i -1}$ by solving the EOM for (\ref{leh}) for 
$A(r)=1$, which obtains the results already given in \cite{Kol:2002xz}. 
It turns out that EOM for the truncated Lagrangian 
(\ref{leh}) ultimately reduces to a single first order differential equation that is given by
\begin{equation}
\frac{d}{d\,r}\Big(r^{d-3}\,(1-A(r))- \frac{2\Lambda}{(d-1)(d-2)} r^{d-1}\Big)=0\,.
\label{eomeh}
\end{equation}
This readily solves to give the solution 
\begin{equation}
A = 1-  \frac{2\Lambda}{(d-1)(d-2)}r^2 - \frac{M}{r^{d-3}}
\end{equation}
and thus we have the static black hole metric as
\begin{eqnarray}\label{construnc4}
ds^2 &=& -\Big(1-  \frac{2\Lambda}{(d-1)(d-2)}r^2 - \frac{M}{r^{d-3}}\Big)\,dt^2 +
\frac{1}{\Big(1-  \frac{2\Lambda}{(d-1)(d-2)}r^2 - \frac{M}{r^{d-3}}\Big)}\,dr^2\nonumber\\
&+& (\frac{d_1-1}{d-3})\, r^2\,d S^2_{(d_1)}+ (\frac{d_2-1}{d-3})\, r^2\,d S^2_{(d_2)},
\end{eqnarray}
with $d_1>1, d_2>1$. Notice that constant coefficient before $dS^2$ indicates a solid angle deficit 
which depends upon dimension of sphere. The metric (\ref{construnc4}) describes a black hole with 
horizon topology $S^{d_1}\times S^{d_2}$.  Note that for $\Lambda =M=0$ spacetime is not Minkowski 
because of solid angle deficits which produce non-zero Riemann curvature as could be seen from the 
Kretschmann scalar, $K=R_{\mu\nu\rho\sigma}R^{\mu\nu\rho\sigma}$  which reads as 
\begin{eqnarray}
K&=&
   \frac{d_1 d_2
   (d-4)
   (d-3)}{2 (d_1-1)
   (d_2-1)}\frac{1}{r^4}+\Big(1+\frac{2}{d-1}-\frac{2}{d-2}\Big)\Lambda^2\nonumber\\
&+&  \frac{1}{4} (d-3)
   (d-2)^2
   (d-1) \frac{M^2}{r^{2(d-1)}}\,.
\end{eqnarray}
Clearly it is non-zero when $\Lambda =M=0$ and spacetime has singularity at $r=0$. However it has 
weaker divergence ($~1/r^4$) as compared to black hole ($1/r^{2(d-1)}$).  When $\Lambda$ and $M$ vanish, 
the solution coincides with the one proposed in \cite{Kol:2002xz}, see also \cite{Emparan:2011ve}, as the 
mediator solution in some topology changing transitions in the space of higher dimensional black hole 
solutions. Note that the black hole potential, $M/r^{d-3}$, remains unaltered by two-spheres topology 
(i.e. $SO(d_1) \times SO(d_2)$ symmetry of the metric). It only rescales $\Lambda$, else it makes no 
difference at all. Thus for Einstein gravity, the black hole solution is neutral to product topology; 
i.e. it doesn't matter whether it is $S^{d_1}\times S^{d_2}$ or simply $S^{d_1+d_2}$.

\subsubsection{Black string and brane}

For $d_1=1$, it turns out that solution cannot accomodate $\Lambda$ because $1$-sphere (circle) has 
no intrinsic curvature. Instead we have the well known solution
\begin{eqnarray}
ds^2 = -\Big(1- \frac{M}{r^{d-3}}\Big)\,dt^2 +
\frac{1}{\Big(1-  \frac{M}{r^{d-3}}\Big)}\,dr^2+ dz^2 + r^2\,d S^2_{(d-3)}\,,
\quad (z\ {\rm periodic}).
\label{construnc5}
\end{eqnarray}
This is the uniform black string solution in which a flat direction is added to a Schwarzschild black 
hole \cite{chr}.

\vspace{4mm}
  
On the other hand if we take one of spheres to be of constant curvature and the other without solid 
angle deficit, then it would give a black brane with the metric, 
\begin{eqnarray}
ds^2 = -A(r)\,dt^2 +
\frac{1}{A(r)}\,dr^2 + r^2\,d S^2_{(d_1)}\pm \frac{1}{k}\,d \Sigma^2_{(d_2)},
\label{construnc6}
\end{eqnarray}
where
\begin{eqnarray}
A(r)= 1- \frac{({d_2}-1)
    }{{d_1}+1}\,k\,r^2-\frac{M}{r^{d_1-1}}\,,\, \, \Lambda=\frac{1}{2} ({d_2}-1)({d_1}+{d_2})\,k.
\end{eqnarray}
Here $\Sigma$ is a space of constant curvature, sphere ($k>0$) or hyperboloid ($k<0$) or flat ($k=0$). 
That is, a constant curvature space is added to a Schwarzschild-dS/AdS black hole in $d=d_1+2$ 
dimension and hence it may be taken as a uniform black brane \cite{chr}. 

\subsubsection{Generalized Nariai metric}
 
For $M=0$, we have product of two spaces of constant curvature, $R^{d_1+2}\times S^{d_2}$, it is a 
generalized Nariai solution \cite{nar, d3} of $R_{ab}=\Lambda g_{ab}$. Let us consider product of two 
constant curvature spaces, $R^2 \times S^2$. If the curvatures are equal, it is Nariai solution of 
$R_{ab}=\Lambda g_{ab}$, on the other hand if they are equal and opposite in sign, then it is an 
Einstein-Maxwell solution \cite{ber,rob} describing gravitational field of uniform electric field. 
Contrary to general behaviour of such spacetimes, the former is conformally non-flat while the latter 
is conformally flat. Note that both product spaces are of the same dimension, while in our case, 
they are of different dimensions, $R^{d_1+2}\times S^{d_2}$, and that is why we call it generalized 
Nariai metric. For $d_2=d_1=2$ where $R^R\times S^2$, the generalized Nariai metric would take the form, 
\begin{eqnarray}
ds^2 = -(1-\frac{\Lambda}{6} r^2)\,dt^2 + \frac{1}{1-\frac{\Lambda}{6} r^2}\,dr^2 + 
r^2\,dS^2_2 +  \frac{2}{\Lambda}\,d \Sigma^2_{2}.
\end{eqnarray}
It is product of  $4$-dimensional dS and $2$-sphere of constant curvature. In fact one can have 
any $n$-dimensional dS with any $m$-sphere of constant curvature to give a generalized Nariaii metric.
 
Thus the two-spheres ansatz we have considered encompasses black objects like hole, string and brane 
as well as  generalized Nariai solution.  

\section{GB black hole}

We can use the same method above to write down the reduced -consistently truncatied- GB Lagrangian 
in term of the variables $A(r),B(r),C(r),D(r)$ given in (\ref{construnc}) and with the use of equations 
(\ref{riem}). So we start by considering the GB Lagrangian (with cosmological constant)
\begin{equation}
L_{GB}=\sqrt{-g}\Big(- 2 \Lambda + {R}^2 -4 {R_\mu}^\nu{R_\nu}^\mu 
+ {R_{\mu\nu}}^{\rho\sigma}{R_{\rho\sigma}}^{\mu\nu}\Big)
\label{lgaussb}
\end{equation}
and truncate it by implementing the ansatz (\ref{construnc}) into it, analogously to (\ref{leh}). 
Since it is not particulary illuminating, the resulting truncated Lagrangian 
is given in detail in the Appendix. 

We begin with the metric (4),
\begin{equation}
ds^2 = -A(r)\,dt^2 +
\frac{1}{A(r)}\,dr^2 + \frac{r^2}{k_1}\,d S^2_{(d_1)}+ \frac{r^2}{k_2}\,d S^2_{(d_2)},
\label{construnc8}
\end{equation}
with $d_1>1, d_2>1$. It turns out that we are able to find analytic solutions when the two-spheres have 
the same dimension $d_1=d_2=:d_0$. This means $d=d_1+d_2+2=2(d_0+1)$ and $k_1=k_2=k
= \frac{2 d_0-1}{d_0-1}$. Let's define 
\begin{equation}
\Psi(r):= 1-A(r),
\label{defpsi}
\end{equation}
then EOM (from the truncated Lagrangian  (\ref{lgb}) ) again becomes a single first order differential 
equation,
\begin{equation}
\frac{d}{d\,r}\left(
r^{2 d_0-3}\Big( \psi ^2 +\frac{d_0}{(d_0-1)^2 (2 d_0-3)}\Big)-
   \frac{\Lambda  r^{2 d_0+1}}{2 (2 d_0+1)(2 d_0-1)( d_0-1)d_0}\right)=0\,.
\label{eomgb}
\end{equation}
It integrates to give  
\begin{eqnarray}
A(r)=1
\pm\sqrt{-\frac{d_0}{(d_0-1)^2 (2 d_0-3)}+\frac{ r^4 \Lambda }{2(2 d_0+1)(2 d_0-1)( d_0-1)d_0\, }
+\frac{M }{ r^{2 d_0-3}}}\,,
\label{solGB}
\end{eqnarray}
where $M$ is an integration constant proportional to mass of the configuration. It represents a black 
hole in an asymptotically dS spacetime. This is a new black hole solution with two-spheres topology 
in GB gravity. The sign $\pm$ is chosen by requiring the solution to asymptotically go over to 
Schwarzschild-dS spacetime in $d=2(d_0+1)$ dimension. Thus we choose negative sign to describe a black 
hole in this setting.  

It is obvious that the solution cannot admit $M=\Lambda=0$ limit and clearly reality of the metric as 
well as existence of horizons would prescribe a bound on mass of the black hole in relation to $\Lambda$.
That is what we next consider.
 
\subsection{Reality and physical bounds for (\ref{solGB})}
\label{physboundGB} 
Since in  absence of black hole ($M=0$), $\Lambda$ must be positive, and hence we shall take both 
$M$ and $\Lambda$ to be always non-negative. 
For concreteness let's set $d_0=2$ which means we are considering $6$-dimensional black hole solution. 
Further for 
simplicity we define $\tilde\Lambda=\frac{\Lambda}{15}$, we write the solution (\ref{solGB}) for $d_0=2$ as 
\begin{equation}
A(r)=1-\sqrt{-2+\frac{1}{4}\tilde\Lambda r^4+\frac{M}{r}}\,,
\label{case2}
\end{equation}
where $\tilde\Lambda$ and $M$ are taken to be non-negative.

Clearly for reality of the solution the discriminant should be $\geq 0$ as well as $A\geq 0$ for the  
existence of black hole horizon. Both these conditions should hold good simultaneously which means 
\begin{equation}
2\leq h(r)\leq 3\,
\end{equation}
where $h(r) := \frac{1}{4}\tilde\Lambda r^4+\frac{M}{r}$. The lower bound guarentees non-negativity 
of the discriminant while the upper ensures existence of horizons bounding a regular region of 
spacetime. The function $h(r)$ has a single minimum at 
\begin{equation}
r_0=\Big(\frac{M}{\tilde\Lambda}\Big)^\frac{1}{5}\,
\end{equation}
and 
\begin{equation}
h(r_0) = \frac{5}{4} \tilde\Lambda^\frac{1}{5} M^\frac{4}{5}. 
\end{equation}

\begin{figure}
\includegraphics[width=8cm,angle=0]{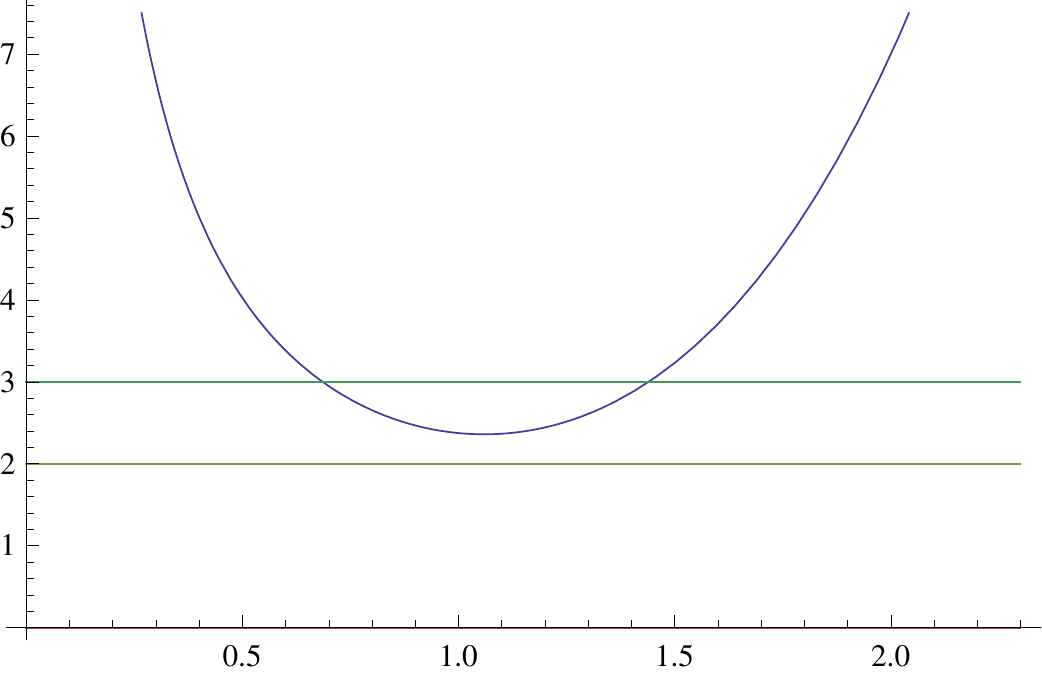}
\caption{Plot of $h(r)$ when $\Big(\frac{8}{5}\Big)^5<  \tilde\Lambda M^4< \Big(\frac{12}{5}\Big)^5$. 
Horizontal lines are $y=3$ and $y=2$, The intersections of $h(r)$ with the upper line $y=3$ define the 
horizons, black hole and cosmological.} \label{region2}
\end{figure}

\begin{figure}
\includegraphics[width=8cm,angle=0]{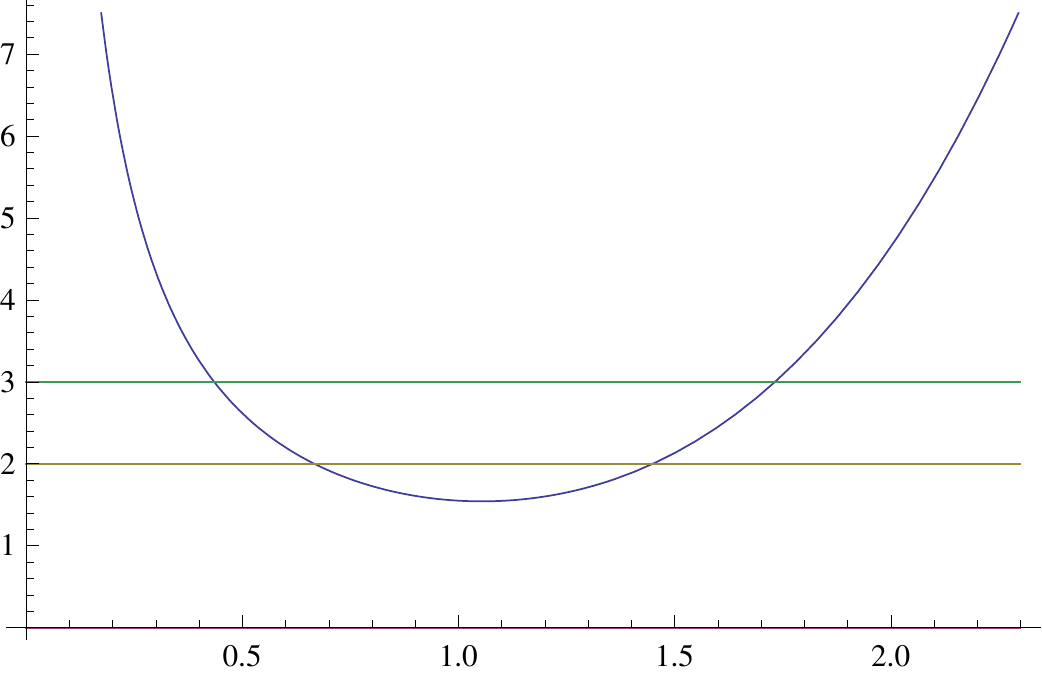}
\caption{Plot of $h(r)$ when $ \tilde\Lambda M^4< \Big(\frac{8}{5}\Big)^5$. Horizontal lines are 
$y=3$ and $y=2$, The intersections of $h(r)$ with the lower line $y=2$ determine the singularities. 
The critical case $ \tilde\Lambda M^4= \Big(\frac{8}{5}\Big)^5$ (not depicted here) means tangency with 
the lower line, and there is no singularity.} 
\label{region3}
\end{figure}

As will be discussed below, for physical viability of black hole we must have $2\leq h(r_0)\leq 3$ 
which implies 
\begin{equation}
2\leq \frac{5}{4} \tilde\Lambda^\frac{1}{5} M^\frac{4}{5}\leq 3\,,
\end{equation}
or, equivalently,
\begin{equation}
\Big(\frac{8}{5}\Big)^5\leq  \tilde\Lambda M^4\leq \Big(\frac{12}{5}\Big)^5\,.
\label{bounds}
\end{equation}
The lower bound is given by the discriminant being non-negative while the upper by existence of 
horizons. The horizons are given by $h(r)=3$ which is a fifth degree equation and can have two 
positive roots giving two horizons, lower ($r_-$) and upper ($r_+$), respectively for black hole 
and cosmological, dS-like, as shown in Fig. \ref{region2}.
 There is an unusual feature of this class of black holes that there occurs a curvature singularity 
 at vanishing of  the  discriminant, $h(r)=2$. As a matter of fact the Ricci scalar for GB black hole 
 (\ref{case2}) is given by 
 $$
 R=\frac{70 M^2+6 M r \left(15 \tilde\Lambda  r^4-56\right)+r^2 \left(3
   \tilde\Lambda  r^4-8\right) \left(5 \tilde\Lambda  r^4-48\right)}{r^4
   \left(\frac{4 M}{r}+\tilde\Lambda r^4-8\right)^{3/2}}\,.
$$
which clearly diverges for $h(r_1)=2$ unless numerator also vanishes at $r_1$. The numerator and 
denominator both vanish simultaneously only for $\tilde\Lambda M^4=\Big(\frac{8}{5}\Big)^5$ at 
$r_1=\frac{5\, M}{8}= \Big(\frac{8}{5\,\tilde\Lambda }\Big)^{\frac{1}{4}}$ making $R$ finite. 
This marks the limiting minimum for black hole mass at which $r_1$ becomes the minimum $r_0$ as shown 
in Fig. \ref{region3}. The remarkable property of this class of black holes is therefore existence of 
extremal value for mass which is a minimum. 

This is in addition to the central singularity at $r=0$. For a black hole, the  latter is always 
covered by a horizon, now the question arises how to manage the former. One of the possibilities 
is to cover it with a horizon but there is no physical source that could produce a horizon to cover it.
There are only $M$ and $\Lambda$ which could only produce the familiar horizons, the former black hole 
horizon covering the central singularity and the latter giving cosmological horizon. The only option 
then left for an acceptable black hole spacetime is therefore not to let it ocuur. The above bounds
(\ref{bounds}) on mass for a given $\Lambda$ precisely do that as demonstrated in Figs ref{region2}, 
\ref{region3}. This requires both $\Lambda,\,M$  to be non-zero. It can be easily seen that either of 
them being zero makes non-central ($h(r)=2$) singularity naked. Not only that it remains naked for 
$\Lambda < 0, \, M>0$ and hence $\Lambda$ must always be positive. This is a new property of this class 
of black holes. In this setting, black hole can thus exist only in asymptotically de Sitter spacetime. 
That is, presence of positive $\Lambda$ is critical for black hole existence. Very recently a similar 
result has also been obtained for stability of pure Lovelock black holes \cite{ganno} where $\Lambda$ 
makes otherwise unstable black hole stable. 

\section{E-GB black hole}
\label{egbbh}
Now we consider the E-GB Lagrangian
\begin{equation}
{L_{E-GB}}=
\sqrt{-g} \Big(-2 \Lambda + \alpha_1{R} + \alpha_2\,({R}^2 -4 {R_\mu}^\nu{R_\nu}^\mu 
+ {R_{\mu\nu}}^{\rho\sigma}{R_{\rho\sigma}}^{\mu\nu})
\Big)\,,
\label{legb}
\end{equation}
with separate parameters $\alpha_1$, $\alpha_2$, so we can recover the GB and EH cases as limits with 
either parameter vanishing. The consistent truncation of (\ref{legb}) under  (\ref{riem})  is given by 
(\ref{leh}) and (\ref{lgb}) (Appendix B).

Again we consider the specific metric, 
\begin{equation}
ds^2 = -A(r)\,dt^2 +
\frac{1}{A(r)}\,dr^2 + \frac{d_0-1}{d-3}r^2\Big(\,d S^2_{(d_0)}+ d S^2_{(d_0)}\Big),
\label{construnc10}
\end{equation}
Now the EOM for the variable $A(r)$ takes the form 
\begin{eqnarray}
&&\frac{d}{d\,r}\left(\alpha_1\, r^{2 d_0-1}\Big(\frac{1}{2(2 d_0^2-3 d_0+1)}\Big) \Psi\right.\nonumber\\ 
&+&\left. \alpha_2\, r^{2 d_0-3}\Big( \Psi ^2 
+\frac{d_0}{(d_0-1)^2 (2 d_0-3)}\Big)-
   \frac{\Lambda  r^{2 d_0+1}}{2(2 d_0+1)(2 d_0-1)( d_0-1)d_0}\right)=0\,,
\label{eomegb}
\end{eqnarray}
where $\Psi(r)$ has been defined in (\ref{defpsi}). This integrates to give the solution
\begin{eqnarray}
 A(r)&=&1+\frac{\alpha_1 r^2}{4 (2 d_0-1)(d_0-1)\alpha_2 }\nonumber\\
&-&\Big(-\frac{d_0}{(d_0-1)^2 (2 d_0-3)}
+\frac{\alpha_1^2 r^4}{4^2 (2 d_0-1)^2(d_0-1)^2 \alpha_2^2}\nonumber\\&+&
\frac{ \Lambda r^4 }{2(2 d_0+1)(2 d_0-1)( d_0-1)d_0\,
\alpha_2 }+\frac{M}{\alpha_2\, r^{2 d_0-3}}\Big)^\frac{1}{2}\,,
\label{solEGB}
\end{eqnarray}
where we have chosen negative sign before the radical for the same reason as for GB case.

In the limit $\alpha_2\to 0$ we recover
\begin{equation}
\lim_{\alpha_2\to 0}A(r)= 1-  \frac{\Lambda }{(2 d_0^2+d_0)\,\alpha_1}r^2-  \frac{2 M ((2 d_0-3)
d_0+1)}{\alpha_1\,r^{2 d_0-1}} \,,
\end{equation}
the corresponding Schwarzschild-dS solution (\ref{construnc4}) for $d_0=\frac{d}{2}-1$, $\alpha_1=1$ 
with an  appropiate redefinition of the mass parameter, $M$\,.

\vspace{4mm}

Let us now set $d_0=2$ and then 
\begin{equation}
A(r)=1+\frac{\alpha_1}{12\, \alpha_2}r^2
-\sqrt{-2+\Big(\frac{ \alpha_1^2}{(12\,\alpha_2)^2}+\frac{ \Lambda}{60\, \alpha_2}\Big)r^4+
\frac{M}{\alpha_2\,r}}
\label{case3}
\end{equation}
It is interesting to compare this solution with the solution with one-sphere topology, 
\begin{equation}
ds^2 = -A(r)\,dt^2 +
\frac{1}{A(r)}\,dr^2 + r^2\,d S^2_{(d-2)},
\label{1sph}
\end{equation}
with 
\begin{equation}
A(r)_{\!{}_{one-sphere}}=1+\frac{\alpha_1}{12\, \alpha_2}r^2
-\sqrt{\Big(\frac{ \alpha_1^2}{(12\,\alpha_2)^2}+\frac{ \Lambda}{60\, \alpha_2}\Big)r^4+\frac{M}{\, 
\alpha_2\,r}}\,.
\label{1sphA}
\end{equation}
It is indeed the same as the above without $-2$ under the radical. Note that the former is not 
asymptotically flat for $\Lambda=0$ while the latter is asymptotically flat, Minkowski. The other 
difference of course is that in the former metric, each sphere has a solid angle deficit which cancel 
out each other to give a $\Lambda$-vacuum spacetime. This is true more generally for $d=2d_0+2$ where 
the former will have $-\frac{d_0}{(d_0-1)^2 (2 d_0-3)}$ under the radical while the latter would be 
free of it.

\subsection{Physical bounds for (\ref{solEGB})}

Let us rewrite (\ref{solEGB}) as
\begin{eqnarray}
A(r)&=&1+D r^2
-\sqrt{f(r)}\,,\nonumber\\
f(r)&=& -C +\frac{1}{4} E^{2 d_0+1} r^4+\frac{1}{2 d_0-3}B^{2 d_0+1}\frac{1 }{r^{2 d_0-3}}\,,
\label{newnot}
\end{eqnarray}
with
\begin{eqnarray}
E^{2 d_0+1}&=& \frac{\alpha_1^2 }{4 (2 d_0-1)^2(d_0-1)^2 \alpha_2^2}
+\frac{\Lambda  }{2(2 d_0+1)(2 d_0-1)( d_0-1)d_0\,\alpha_2 }\nonumber\\
B^{2 d_0+1}&=:&(2 d_0-3)\frac{M}{\alpha_2}\nonumber\\
C&=&\frac{d_0}{(d_0-1)^2 (2 d_0-3)}\nonumber\\
D&=&\frac{\alpha_1 }{4 (2 d_0-1)(d_0-1)\alpha_2 }
\label{definitions}
\end{eqnarray}
Note that since their exponent is odd, the signs of $E$ and $B$ are those of their respective right 
hand side in the definition.

\vspace{4mm}

Let us consider the case $\alpha_1>0,\,\alpha_2>0$ (the same sign for the $EH$ and $GB$ coefficients 
is required for the theory to be ghost free\cite{BD}) as well as $\Lambda>0$ and $M>0$, 
which imply $E>0,\,B>0$. Note from (\ref{definitions}) that there is a window for negative $\Lambda$ 
and still keeping $E>0$.

The minimum for $f$ is attained at $\displaystyle r_0= \frac{B}{E}$. 
Following on the same lines as before (see section (\ref{physboundGB}), we obtain the bounds as follows: 
\begin{equation}
C\,\leq\,\frac{2 d_0+1}{4(2 d_0 -3)}E^{2 d_0-3} B^4\,\leq\,C+(1+D(\frac{B}{E})^2)^2 \,
\label{generalbound}
\end{equation}
where $C$, given in (\ref{definitions}), is a positive quantity determined by spacetime dimension. 
Since $E, B, D$ are positive, it is clear, for given $\alpha_1,\alpha_2$, there exists a range of 
values for $E$ and $B$ (i.e. for $\Lambda$ and $M$) fullfilling
the bounds (\ref{generalbound}). Thus as before non-central curvature singularity at the vanishing 
of the radical could be avoided by suitable prescription on black hole mass for given $\Lambda$. 

\section{Thermodynamics of black holes}

We will use the notation of Eq (\ref{definitions}) and of course we assume that the conditions given 
in Eq (\ref{generalbound}) hold true, which guarentee existence of horizons. Let's denote black hole 
horizon by $r_h$. Of course $r_h$ and $M$ could be traded for each other, just by requiring the 
function $M(r_h)$ to keep $A(r_h)=0$ while varying $r_h$. The entropy is calculated from the First 
Law of thermodynamics by following the standard procedure.

We write the identity
\begin{equation}
A(r,M(r))=0 
\label{identit}
\end{equation}
as an implicit equation for the function $M(r)$ introduced above, therefore we have the identity 
\begin{equation}
A'(r,M(r)) + \frac{\partial A}{\partial M}|_{A=0}\,M'(r)=0\,,
\label{deridentit}
\end{equation}
where $A'(r,M(r))$ denotes derivative relative to first argument $r$. Employing the Eucidean method 
(with periodic time to eliminate a conical singularity), we identify $A'(r,M(r))$ as the Hawking 
temperature $\displaystyle T(r) =\frac{A'(r,M(r))}{4\pi}$. 

Thus we have
\begin{equation}
\frac{M'(r)}{T} = - 4\pi \frac{1}{\frac{\partial A}{\partial M}|_{A=0}}
= 8\pi \alpha_2(1+D r^2) r^{2d_0-3}\,,
\label{mprima}
\end{equation}
and we can compute the entropy by integrating the First Law
\begin{equation}
S =\int\frac{d M}{T} = \int\frac{M'(r)}{T(r)} dr = 8\pi\alpha_2\int(1+D r^2) r^{2d_0-3} dr
= 8\pi\alpha_2\, r_h^{2 d_0}\Big(\frac{1}{(2 d_0-2)r_h^2} + 
\frac{D }{2 d_0}\Big)\,.
\label{entr}
\end{equation}
where we have assumed that the entropy vanishes when the horizon shrinks to zero.

Of course the parameter $M$ used in our derivation is identified with the mass except for an 
overall factor that will depend on the dimension of the spacetime and linearly on area of two-spheres 
at unit radius. With this factor 
installed we get
\begin{equation}
S \simeq \,{\bf A} \times \Big(\frac{\alpha_2}{(2 d_0-2)r_h^2} + \alpha_2
\frac{D }{2 d_0}\Big) = {\bf A}\times \Big(\frac{\alpha_2}{(2 d_0-2)r_h^2}+ 
\frac{\alpha_1}{8 d_0 (2 d_0-1)(d_0-1)}\Big)= 
{\bf A}\times \Big(\frac{\bar\alpha_2}{r_h^2}+ 
\bar\alpha_1 \Big)\,,
 \label{entr2}
\end{equation}
where $\bf A$ denotes horizon area and $\bar\alpha_1 = \alpha_1/(8d_0(2d_0-1)(d_0-1))$ and 
$\bar\alpha_2 = \alpha_2/(2d_0-2)$.
\vspace{4mm}

The temperature, in terms of $r_h$, is
\begin{equation}
T =\frac{1}{ 8\pi\,(1+D r_h^2)r_h}\Big( (1+D r_h^2)  (2 d_0-3 +(2 d_0+1)D r_h^2) +(2 d_0-3)C 
-\frac{2 d_0+1}{4}E^{2 d_0+1} r_h^4               \Big)\,.   
\label{temp}
\end{equation}
For instance, for the pure GB case ($\alpha_2=1, \alpha_1=0 \,(\Rightarrow D=0)$) and $d_0=2$ ($d=6$), 
we obtain
\begin{equation}
T= \frac{1}{2\pi }(-\frac{3}{r_h}+\frac{5}{4}\frac{M}{r_h^{2}} )\,,\qquad\qquad 
S\simeq \frac{ {\bf A}}{ r_h^2} \simeq  r_h^2 \simeq {\bf A}^{\frac{1}{2}} \,.
\label{temp6}
\end{equation}

It is worth noting that these parameters bear the same universal relation to $r_h$ as established 
in \cite{dpp1} for pure Lovelock one-sphere topology in the critical dimension $d=2N+2$, 
here for $N=2$ case. In particular, for pure GB black hole, 
$T= \frac{1}{2\pi }(-\frac{1}{r_h}+\frac{5}{4}\frac{M}{r_h^{2}} )$ and $S=4\pi r_h^2$. 
Thus thermodynamics parameters do not distinguish between one or two-sphere topology, 
save for numerical factors. 

\vspace{4mm}

Let us finally discuss the stability of the GB black hole. Here we take $\alpha_1=0$. $\alpha_2=1$.
The temperature obtained above gives, for the GB case ($D=0$)

\begin{equation}
T =\frac{1}{8\pi\, r_h}\Big(   (2 d_0-3)(C +1) 
-\frac{2 d_0+1}{4}E^{2 d_0+1} r_h^4                  \Big)\,,  
\label{tempGB}
\end{equation}
and for it to be positive we need 
\begin{equation}\frac{2 d_0+1}{4(2 d_0-3)}E^{2 d_0+1} r_h^4 \leq (C +1)\,.
\label{newb}
\end{equation}
Actually this requirement is more restrictive than the bound determined before (\ref{generalbound}), 
which written in terms of 
$r_0$ (the minimum of $f$) becomes, for the side we are interested in, 
$\displaystyle \frac{2 d_0+1}{4(2 d_0-3)}E^{2 d_0+1} r_0^4 \leq (C +1)$. Since 
according to our construction $r_0<r_h$, in order to keep $T$ positive, we must replace the rhs of the 
bound (\ref{generalbound}) by (\ref{newb}), in the $D$=0 case.

Once we set the bounds to have positive temperature, local thermodinamical stability will correspond to 
positive specific heat, $\displaystyle C_e=\frac{d M}{d T}$. Using the expressions $M(r_h)$ and $T(r_h)$ 
we have
$$ C_e = \frac{M'(r_h)}{T'(r_h)} = \frac{M'(r_h)}{T}\frac{T}{T'(r_h)}\,,
$$
and we have from (\ref{mprima}), $\frac{M'(r)}{T} 
= 8\pi  r^{2d_0-3} >0$,. Since $T>0$ we need $T'(r_h)>0$ for $C_e$ to be positive. But clearly
we infer from (\ref{tempGB}) that $T'(r_h)<0$ and thus our solution is not thermodinamically stable. 
 Since the case of one sphere (instead of two) could be obtained by setting the parameter $C$ to zero 
 in Eqs  (\ref{definitions}) (see (\ref{case3}), (\ref{1sphA})), it is clear that the 
instability found here is the same as in the spherically symmetric case.

\section{Solutions with two-spheres of constant curvature}

In this section we apply the same consistent truncation method to obtain new solutions 
to EH, GB and E-GB with two-spheres of constant curvature. The metric is therefore of the form 
\begin{equation}
ds^2 = -A(r)\,dt^2 +
\frac{1}{A(r)}\,dr^2 + \frac{d_1-1}{k}\,d S^2_{(d_1)}+ \frac{d_2-1}{k}\,d S^2_{(d_2)},
\label{construnc2}
\end{equation}
with $d_1>1, d_2>1$ and $k>0$. Note that constant curvature of each sphere, $\frac{k}{d_i-1}$, is 
different according to its dimensionality. We are seeking solution of $\Lambda$-vacuum equation and 
hence $(t, r)$ space has also to be of constant curvature. That is, it can be $dS_2$ or $AdS_2$. The 
topology is therefore $dS_2/AdS_2\times S^{d_1}\times S^{d_2}$. This is a generalized Nariai spacetime 
\cite{nar}. \\



Clearly $A=1-kr^2$ in the above metric is the Einstein solution and $\Lambda$ is determined as $\Lambda = \frac{1}{2}(d_1+d_2)k >0$. 

When $k<0$ there are are hyperboloids in place of spheres, and the solution could then be written as, 

\begin{eqnarray}
ds^2 = -(1+ |k| r^2)\,dt^2 +
\frac{1}{1+ |k| r^2}\,dr^2 + \frac{d_1-1}{|k|}\,d H^2_{(d_1)}+ \frac{d_2-1}{|k|}\,d H^2_{(d_2)},
\label{construnc3}
\end{eqnarray}

with $\Lambda=\frac{1}{2} (d_1+d_2) k<0$ and topology, $AdS_2\times H^{d_1}\times H^{d_2}$\,. 
This is a generalized anti-Nariai metric \cite{prod}.  The case $k\to 0$ is Minkowski spacetime (the spheres or hyperboloids acquire infinte radius and become flat), though the metric (\ref{construnc3}) is no longer convenient to describe such a limit.\\
For the case of GB, we find solutions for (\ref{construnc2}) with $A=1-k r^2$ and $k$ 
determined below.
Curvatures $k_1$ and $k_2$ are constrained by a third degree polynomial equation,
\begin{eqnarray}
&&({d_1}-1) ({d_2}-1) {k_1}^2
   {k_2} \left(({d_1}-2)^2
   ({d_1}+3)-2 (({d_1}-3)
   {d_1}+3)
   {d_2}\right)\nonumber\\
&+&({d_1}-1)
   ({d_2}-1) {k_1} {k_2}^2
   \left(2 {d_1} (({d_2}-3)
   {d_2}+3)-({d_2}-2)^2
   ({d_2}+3)\right)\nonumber\\&+&(3-{d_1})
   ({d_1}-2) ({d_1}-1)^2 {d_1}
   {k_1}^3+({d_2}-3) ({d_2}-2)
   ({d_2}-1)^2 {d_2} {k_2}^3=0\,,
\end{eqnarray}
with $\Lambda$ (here $\Lambda$ includes a dimensional factor originated by the dimensionality of the 
GB Lagrangian) is given by
\begin{eqnarray}
\Lambda &=&\frac{1}{8} \Big(2 ({d_1}-1) {d_1}
   ({d_2}-1) {d_2} {k_1}
   {k_2}+({d_1}-3) ({d_1}-2)
   ({d_1}-1) {d_1}
   {k_1}^2\nonumber\\ &+&({d_2}-3) ({d_2}-2)
   ({d_2}-1) {d_2} {k_2}^2\Big)\,,
\end{eqnarray}
and 
\begin{equation}
k = \frac{({d_1}-1) {k_1}
   (({d_1}-3) ({d_1}-2)
   {k_1}+({d_2}-1) {d_2}
   {k_2})}{({d_1}-2) ({d_1}-1)
   {k_1}+({d_2}-1) {d_2}
   {k_2}}\,.
\end{equation}

For the equal dimension spheres, $d_1=d_2=:d_0$ and $k_1=k_2$, they become 
\begin{equation}
\Lambda=\frac{1}{2} (d_0-1) d_0 ((d_0-3) d_0+3) k^2\,,\qquad k=\frac{((d_0-3)
   d_0+3) }{d_0-1} \, k_1.
\end{equation}

We continue with (\ref{construnc2}) and $A=1-k r^2$. As expected for E-GB $k_1$ and $k_2$ 
have also to satisfy a third order polynomial equation. 
We consider the simple case of $d_1=d_2=d_0$ and $k_1=k_2$, and then we obtain 
 
\begin{equation}
\Lambda = \frac{1}{2} (d_0-1) d_0
   k_1 (2\alpha_1+4 ((d_0-3) d_0+3) k_1\alpha_2
   )
\end{equation}
and
\begin{equation}
k = \frac{(d_0-1)  (\alpha_1+4 ((d_0-3) d_0+3) k_1
   \alpha_2)}{\alpha_1+4 (d_0-1)^2 k_1 \alpha_2} k_1 ,
\end{equation}
which reduces to EH solution for $\alpha_2\to 0$ and to GB for $\alpha_1\to 0$. \\ 

These GB and E-GB cases also admit hyperboloids in place of spheres. The constant spheres are 

 spacetimes whereas constant hyperboloids are generalized anti-Nariai spacetimes.

\section{Discussion}

It is well-known that vacuum equation for Einstein as well as for general Lovelock gravity in spherical 
symmetry ultimately reduces to a single first order equation which is an exact 
differential \cite{Wheeler:1985qd,Wheeler:1985nh,Whitt:1988ax,Camanho:2011rj,Maeda:2011ii,dpp} and hence 
can be integrated trivially. As a matter of fact, 
the equation then turns purely algebraic for one-sphere topology with $SO(d-2)$ symetry. 
Interestingly it turns out that this feature is carried through even for two-spheres topology 
with $SO(d_0)\times SO(d_0)$ symmetry where $d=2(d_0+1)$. In particular the equations,   
(\ref{eomeh}), (\ref{eomgb}) and (\ref{eomegb}), refer respectively to Einstein, 
GB and E-GB gravity which yield static black hole solutions. This result obviously raises
the question as to whether this feature is also carried over to Lovelock gravity in general. 
The answer is in affirmative and it would be taken up separately in a forthcoming 
paper \cite{nj14-2}. 

\vspace{4mm}

For black string, there occurs local topology change as horizon radius increases from that of a 
black hole $S^{d_0}$ to  of black string $S^{d_0-1} \times S^1$. This change is negotiated 
through \cite{Kol:2002xz} a Ricci flat space over a double cone formed by two spheres with solid angle 
defecit. The distinguishing feature of this class of Dotti-Gleiser black holes \cite{Dotti1} is that 
horizon space is an Einstein space with both  Weyl and Riemann  curvatures being covariantly 
constant. Contrary to what is said in the literature following Ref. \cite{Dotti1},  space is though 
not maximally symmetric; i.e. Riemann curvature is not given in terms of the metric but its covariant 
derivative is zero. 

This non-zero Weyl curvature makes 
 non-trivial contribution in the black hole potential which gives rise to non-central naked 
 singularity. For Einstein black hole, the topology is $S^{d_1} \times S^{d_2}$ while for GB and 
 E-GB it is $S^{d_0} \times S^{d_0}$, and further it makes no contribution to potential for the 
 former. This is because for Einstein Riemann and Weyl curvatures do not enter into the 
 equation while they do so for Lovelock.

\vspace{4mm}

For GB and E-GB black holes, what $SO(n)\times SO(n)$ symmetry entails is occurrence of an additional  non-central curvature singularity which could be let not to occur by suitable prescription on black hole mass for a given $\Lambda > 0$ (for E-GB black hole, a narrow window of negative $\Lambda$ is also permitted). The two extremal limits for mass are defined by non-occurrence of non-central naked singularity 
(intersection with lower line in Fig. \ref{region3}) and existence of horizons 
(intersection with upper line in Fig. \ref{region2}). The range for mass is given in Eqs (23) and (33) which ensures absence of naked singularity for GB and E-GB black hole in dS spacetime. Also non-central singularity cannot be avoided when $M=0$ or $\Lambda\leq 0$.  
Thus $\Lambda$ plays a very critical role for existence of this class of black holes. This reminds one of the recently obtained result in which $\Lambda$ makes 
otherwise unstable pure Lovelock black hole stable by similarly prescribing a range of values for mass \cite{ganno}.

\vspace{4mm}

Further it turns out that black hole thermodynamics does not however distinguish between two-spheres 
and one-sphere topolgy as expressions for temperature and entropy of black hole remain essentially 
the same. For pure Lovelock black hole with spherical symmetry thermodynamics is universal; i.e. 
temperature and entropy bear the same relation to horizon radius in all odd ($d=2N+1$) and even 
($d=2N+2$) dimensions where $N$ is the degree of Lovelock Lagrangian \cite{dpp1}. It is interesting 
that this universality continues to hold true even for two-spheres topology black holes in GB and 
E-GB gravity. 

\vspace{4mm}

Finally, let us mention that the technique of the truncated Lagrangians allows us to find new 
solutions with two-spheres of constant curvature (or hyperboloids). They correspond to generalized 
Nariai (or anti-Nariai for hyperboloids) solutions.

\section*{Acknowledgments}
We thank Sushant Gosh for collaboration at an earlier stage of the project. JMP thanks Roberto Emparan 
and Jos\'e Edelstein for helpful conversations. ND thanks Xian Camanho for enlightening and insightful discussion on constant curvature spaces.  JMP acknowledges partial financial support from 
projects FP2010-20807-C02-01,2009SGR502 and CPAN Consolider CSD 2007-00042.

\section*{Appendix. The truncated GB Lagrangian}

After some combinatorics, the GB Lagrangian (\ref{lgaussb}) (without the cosmological constant), 
reduced under (\ref{construnc}), becomes, with the notation introduced in (\ref{riem}),
\begin{eqnarray}
&& L_{GB}=\sqrt{-g}\Big(8\, d_1 d_2 L(0,1) 
   L(a,a')+4\, (d_1-1) d_1
   L(0,1) L(a,b)\nonumber\\&+&8\, (d_1-1)
   d_1 d_2 L(a,a')
  \Big (L(0,a)+L(1,a)\Big)
   +4\, d_1
   (d_2-1) d_2 L(a',b')
  \Big (L(0,a)+L(1,a)\Big)\nonumber\\
   &+&8\, d_1 d_2
   L(0,a) 
   L(1,a')+4\, (d_1-2) (d_1-1)
   d_1 L(a,b)
   \Big(L(0,a)+L(1,a)\Big)\nonumber\\&+& 8\, (d_1-1)
   d_1 L(0,a) L(1,a)+8\,
   d_1 d_2 L(0,a') 
   L(1,a)+8\, d_1 (d_2-1)
   d_2 L(a,a')
  \Big(L(0,a')+L(1,a')\Big)\nonumber\\&+&4\, (d_1-1)
   d_1 d_2 L(a,b)
  \Big (L(0,a')+L(1,a')\Big)+4\, (d_1-1)
   d_1 (d_2-1) d_2
   L(a,a')^2\nonumber\\&+&4\, (d_1-2) (d_1-1)
   d_1 d_2 L(a,a')
   L(a,b)+4\, d_1 (d_2-2)
   (d_2-1) d_2 L(a,a')
   L(a',b')\nonumber\\&+&2\, (d_1-1)
   d_1 (d_2-1) d_2
   L(a,b) L(a',b')+
   (d_1-3) (d_1-2) (d_1-1)
   d_1 L(a,b)^2\nonumber\\&+& 4\, 
   (d_2-1) d_2 L(0,1)
   L(a',b')+4\, (d_2-2) (d_2-1)
   d_2 L(a',b')
   \Big(L(0,a')+L(1,a')\Big)\nonumber\\&+&8 \,
   (d_2-1) d_2 L(0,a')
   L(1,a')+ (d_2-3)
   (d_2-2) (d_2-1) d_2
   L(a',b')^2\Big)\,.
   \label{lgb}
\end{eqnarray}

\end{document}